\documentclass[12pt, preprint]{aastex}
\shorttitle{Spitzer/IRAC Imaging of the AGB Stars in WLM}
\shortauthors{Jackson et al.}

\begin{document}

\def\HI{\ion{H}{1}}
\def\HII{\ion{H}{2}}
\def\msun{M_$\sun$}
\def\spitzer{$\it Spitzer$}
\def\micron{$\mu$m}
\def\solar{M$_\sun$}
\def\solaryr{M$_\sun$ yr$^{-1}$}
\def\ml{$\dot{M}$}

\title{A Spitzer/IRAC Census of the Asymptotic Giant Branch
Populations in Local Group Dwarfs. I.  WLM} \author{Dale C. Jackson,
Evan D. Skillman, Robert D. Gehrz, Elisha Polomski, and Charles
E. Woodward} 
\affil{Astronomy Department, University of Minnesota, 116
Church St. S.E., Minneapolis, MN 55419} 
\email{djackson@astro.umn.edu,
skillman@astro.umn.edu, gehrz@astro.umn.edu, elwood@astro.umn.edu,
chelsea@astro.umn.edu}


\begin{abstract} 
We present {\it Spitzer}/IRAC observations at 3.6 and 4.5 \micron\
along with optical data from the Local Group Galaxies Survey to
investigate the evolved stellar population of the Local Group dwarf
irregular galaxy WLM. These observations provide a nearly complete
census of the asymptotic giant branch (AGB) stars. We find 39\% of the
infrared-detected AGB stars are not detected in the optical data, even
though our 50\% completeness limit is three magnitudes fainter than
the red giant branch tip. An additional 4\% of the infrared-detected
AGBs are misidentified in the optical, presumably due to reddening by
circumstellar dust. We also compare our results with those of a
narrow-band optical carbon star survey of WLM, and find the latter
study sensitive to only 18\% of the total AGB population. We detect
objects with infrared fluxes consistent with them being mass-losing
AGB stars, and derive a present day total mass-loss rate from the AGB
stars of \ml~=~0.7-2.4$\times$10$^{-3}$ \solaryr . The distribution of
mass-loss rates and bolometric luminosities of AGBs and red
supergiants are very similar to those in the LMC and SMC and the
empirical maximum mass-loss rate observed in the LMC and SMC is in
excellent agreement with our WLM data.
\end{abstract}



\section{Introduction}
Dwarf galaxies are the most common type of galaxies in the universe.
Understanding their formation and evolution is an important step in
understanding the evolution of all galaxies since they are likely the
building blocks of more massive systems like the Milky Way. Because
the basic properties of dwarf galaxies (luminosity, stellar and
gaseous mass, star formation rate (SFR), and metallicity) are very
different from those of massive galaxies, detailed studies of the
former objects provide us with an important tool to understand better
how the astrophysics of star formation and galaxy evolution depend on
these characteristics. The dwarf galaxies of the Local Group, in
particular, provide a wealth of information about their evolution;
unlike more distant objects, we are able to resolve them into
individual stars to study the characteristics of their stellar
populations and to create detailed star formation histories (SFHs).

The Local Group dwarf irregular (dI) galaxy WLM, first discovered by
\citet{wol09} and independently rediscovered by Lundmark and Melotte
\citep{mel26}, has been the subject of many investigations over the
past two decades \citep[cf.,][]{van00}. \citet{huc81} and
\citet{huc86} first addressed the \HI\ properties of the galaxy,
finding it to be gas-rich, in keeping with its dI
morphology. \citet{jac04} were the first to resolve the galaxy in \HI
, and found a double-peaked flux distribution.  Because WLM is
relatively nearby \citep[m$-$M~=~24.88;][]{dol00}, the stellar
population is easily resolved, enabling high quality optical
photometry. \citet{dol00} and \citet{rej00} find WLM's SFH suggests
the bulk of the stars formed over 9 Gyr ago. This episode of star
formation was followed by a gradual decrease in the star formation
rate until the most recent event that began between 1 and 2.5 Gyr ago
and continues today \citep{hod95,ski89,hun93,lee05}. \citet{jac06}
studied the emission from hot dust and polycyclic aromatic
hydrocarbons (PAHs) and found only very low surface brightness
emission coincident with the highest surface brightness \HII\ regions,
as is expected from a galaxy with relatively low metallicity
\citep[12~+~log(O/H)~=~7.83;][]{lee05} and current star formation rate
\citep[0.003 M$_\sun$ yr$^{-1}$;][]{mat98}.

The evolved stellar population of WLM was examined by \citet{coo86}
and \citet{bat04}, who both found an extremely high carbon star to
M-type asymptotic giant branch (AGB) star ratio in accordance with its
low metallicity \citep{ibe83}. Other studies using narrow-band filters
centered on the CN and TiO absorption features have enabled the study
of AGBs in a limited sample of Local Group galaxies
\citep{coo86,now01,bat05}. Because optical studies can suffer
significantly from extinction effects (due to both the host galaxy
interstellar medium (ISM) and the winds of individual stars), many AGB
studies have been performed in the infrared
(IR)\citep[e.g.,][]{boy06,zij96,woo98,van99,van05}. The majority of
this effort has been limited to the Galaxy and the Magellanic Clouds
because a combination of poor angular resolution and observational
sensitivity made similar investigations in more distant systems
impossible. However, near-IR studies of other Local Group galaxies
(particularly dwarf spheroidals) recently have been conducted with the
large-aperture ground based telescopes
\citep{bab05,cio05,dav05,kan05,soh06}.

In this paper we present \textit{Spitzer Space Telescope} imaging of
WLM at 3.6 and 4.5 \micron . These observations provide us with the
first high-resolution (0.9\arcsec\ pixel$^{-1}$~=~4 pc pixel$^{-1}$)
high-sensitivity thermal-IR images of this galaxy, enabling us to
investigate the evolved stellar population in WLM. Section
\ref{observations} describes the observations and reduction of the
\spitzer\ data as well as previously unpublished data from the Local
Group Survey \citep{mas06}. In \S \ref{morphology} we discuss the
morphology of WLM as a function of wavelength. We address both the
optical and IR photometry in \S \ref{photometry} and discuss the AGB
and carbon star population. We derive mass-loss rates and wind optical
depths in \S \ref{mass_loss}. In \S \ref{luminosity} we investigate
the contributions from various stellar types to the near-IR luminosity
of WLM.


\section{Observations and Data Reduction}\label{observations}
\subsection{Infrared and Optical Data and Photometry}
\spitzer\ Space Telescope \citep{wer04} data of WLM were obtained with
the Infrared Array Camera \citep[IRAC;][]{faz04} in two separate AORs
on 2003 December 12 UT and 2005 December 23 UT (AORKEYs: r5051136 and
r15892480, respectively). Both AORs of WLM were done as part of a
larger guaranteed time observing program (PID: 128, PI: R.D. Gehrz).

\begin{figure}
\epsscale{0.5}  
\plotone{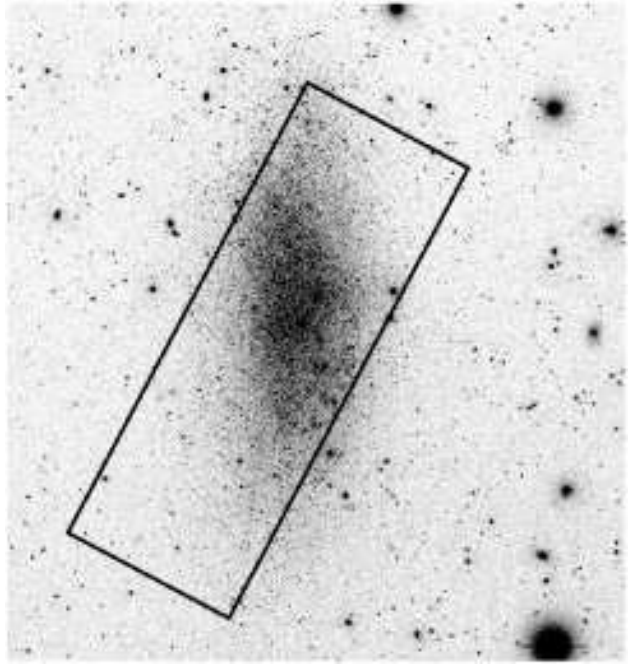}
\caption{\label{coverage}LGGS I-band image of WLM from \citet{mas06}. The region
with coverage in all four IRAC bands is overlayed in black. North is
up and East is left.}
\end{figure}

IRAC provides simultaneous imaging at 3.6, 4.5, 5.8, and 8.0 \micron ;
however, the field of view at 3.6 and 5.8 \micron\ are offset from
that at 4.5 and 8.0 \micron .  To map WLM most efficiently a
1$\times$4 mosaic was created, leaving a coverage area in all four
IRAC bands of approximately 4$\arcmin$ $\times$ 12$\arcmin$ centered
on $\alpha$=00:02:00, $\delta$=$-$15:28:30 (Figure
\ref{coverage}). Throughout this paper all discussions refer to this
region only. A total exposure time of 500 seconds per pointing was
used for all bands.

The IRAC data (pipeline version 13.2.0) were reduced using the
  MOPEX\footnote{MOPEX is available from the Spitzer Science Center at
  http://ssc.spitzer.caltech.edu/postbcd/} reduction package, version
  2006 March 1. The overlap routine was used to match the backgrounds
  of individual frames to create a smooth background in the final
  mosaic. Outlier detection, image interpolation, and image
  co-addition were done with the mosaic program which also created the
  final mosaics. The 8.0 \micron\ data taken in 2003 were strongly
  affected by persistent images resulting from bright sources imaged
  in the program immediately before our AORs were executed.  Subtracting
  a median-combined image of all of the dithered frames from each
  frame proved very effective at removing these artifacts.

Point source photometry was done with the DAOPHOT II photometry
package \citep{ste87}. For the relative photometric errors, we adopt
the formal errors produced by DAOPHOT. IRAC has an estimated absolute
calibration accuracy of 3\% \citep{rea05}. Due to the high sensitivity
of these observations, a significant number of background galaxies are
detected. A sharpness clip, which identifies detections that are best
fit by a point spread function (PSF) much broader or narrower than our
template PSF, was applied to all bands to produce our final photometry
list, which proved effective at removing extended objects.

Broadband U, V, and I images of WLM were obtained from the Local Group
Galaxies Survey\footnote{LGGS data are publicly available at
http://www.lowell.edu/$\sim$massey/lgsurvey/}
(LGGS, \citealt{mas06}). The ground-based optical imaging from the LGGS
complements the \spitzer /IRAC imaging nicely because they have
similar angular resolutions. The U image was used only to investigate
the change in galaxy morphology with wavelength and no photometry was
performed on this image. Because our use of the V and I data was
solely to identify stellar types via color magnitude diagrams to then
cross-identify these sources in the near-IR, precise absolute
photometry was not required. Consequently, `stacked' images that have
relative photometric errors of $\sim$10\% were used rather than the
`photometric' images since image processing, including co-addition and
astrometric calibration, were already performed on the former.

No photometric calibrations have yet been released for the V and I
data. Consequently, to calibrate the magnitude offsets between
instrumental and true magnitudes, we compared our photometry to that
of the photometrically calibrated Hubble Space Telescope (HST) Local
Group Stellar Photometry Archive \citep{hol06}. These data are
publicly available, and one HST/WFPC2 pointing ($<$~6 arcmin$^2$) is
entirely contained in our IRAC coverage. The \citet{mas06} data cover
a 36$\arcmin$$\times$36$\arcmin$ field of view, including our entire
Spitzer coverage area. The magnitude offsets and color terms for
sources with centroids matching those in the HST photometry were
calculated using least squares fitting. The appropriate offsets were
then applied to all the detected sources in the V and I
photometry. The optical and IR photometric catalogs were combined by
matching point sources that have the same centroids within a tolerance
of $\pm$1$\arcsec$. Throughout this paper all magnitudes are stated with
respect to $\alpha$~Lyr (Vega).

\subsection{Foreground Star Contamination}
In \S \ref{photometry} we examine the properties of the evolved
stellar population in WLM. Because these objects can have similar
apparent magnitudes and colors to Galactic late-type dwarfs, it is
important to determine the extent to which our data will be affected
by contamination by Galactic foreground stars. We estimated the number
of foreground stars in our IR data using the Milky Way stellar
population synthesis model of \citet{rob03}. We chose a field one
square degree in size centered on the Galactic coordinates of WLM to
provide robust statistics. This model provides magnitudes for modeled
stars in L band, whose central wavelength is close to that of the 3.6
\micron\ IRAC band.

From this model we expect only 9 foreground stars in the area from
which our IR CMD was constructed with $-$13~$<$~M$_{3.6}$~$<$~$-$5. These are
all expected to have [3.6]$-$[4.5] colors very near zero.


\section{Optical and IR Galaxy Morphology}\label{morphology}
\begin{figure}
\epsscale{1}  
\plotone{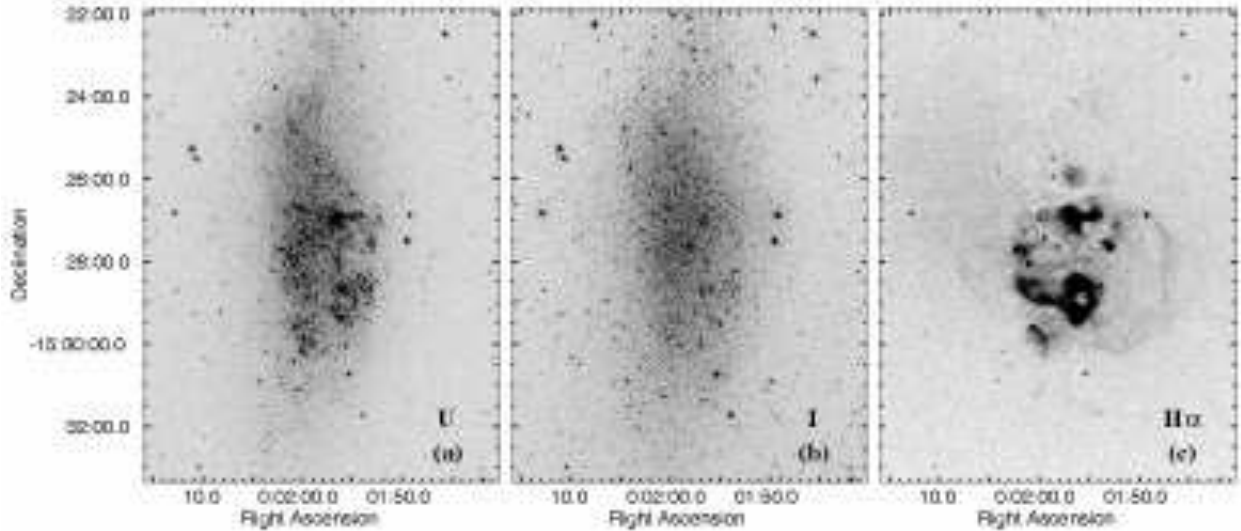}  
\caption{\label{3color}(a) U, (b) I, and (c) continuum subtracted
H$\alpha$ images of WLM from the LGGS \citep{mas06}, showing the
change in morphology with wavelength. Note the concentration of recent
star formation in the south. }
\end{figure}

WLM is a highly-inclined \citep[$\it{i}$ = 69$\degr$;][]{abl77,jac04}
disk galaxy, elongated almost exactly north-south. Its overall
appearance varies considerably from optical to IR wavelengths. Figure
\ref{3color} shows the U (a), I (b), and continuum-subtracted
H$\alpha$ (c) images of WLM. In the U image there is a prominent
feature arcing to the north-west. The rest of the emission consists of
four associations in the central and southern galaxy that
are currently undergoing star formation, as is evident from the
H$\alpha$ emission. In the I image we see the galaxy morphology change
from that of a well defined arc and distinct star forming regions to a
much smoother stellar distribution. The H$\alpha$ emission is confined
to the central and southern parts of the galaxy, with no detected
emission in the northern arc.  There is a large loop of H$\alpha$
emission to the southwest, directly across from which (to the
northeast) is a large area of very low surface brightness H$\alpha$
emission. This low surface brightness feature is coincident with the
central region of WLM that is \HI\ deficient (northeast of the two
\HI\ peaks in \citealt{jac04}). We take this as further evidence of a
partial blowout of the ISM by the massive young stellar population.

\begin{figure}
\plotone{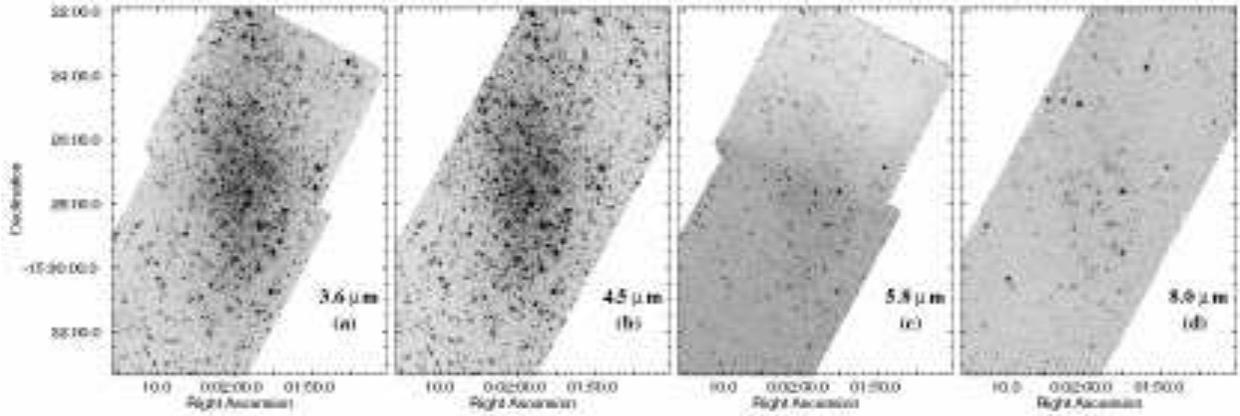}  
\caption{\label{ch12}IRAC 3.6 \micron\ (a), 4.5 \micron\ (b), 5.8
\micron\ (c), and 8.0 \micron\ (d) images of WLM. Due to both
decreased sensitivity and fainter stellar emission at longer
wavelengths we detect significantly few objects than in the short
wavelength bands.}
\end{figure}

The IRAC images of WLM are shown in Figure \ref{ch12}. The appearance
of the galaxy retains the smooth stellar distribution observed in the
I image with the addition of a large population of very luminous
objects throughout the galaxy (AGBs or red supergiants
[RSGs]). \citet{jac06} detected extremely faint diffuse 8.0
\micron\ emission in WLM, coincident with the high surface brightness
\HII\ regions HM7 and HM9 \citep{hod95}.  In \S \ref{photometry}
we discuss how the distribution of AGB stars compares with that of the
other stellar populations.


\section{Comparative Optical and IR Photometry}\label{photometry}
\begin{figure}
\epsscale{0.75}  
\plotone{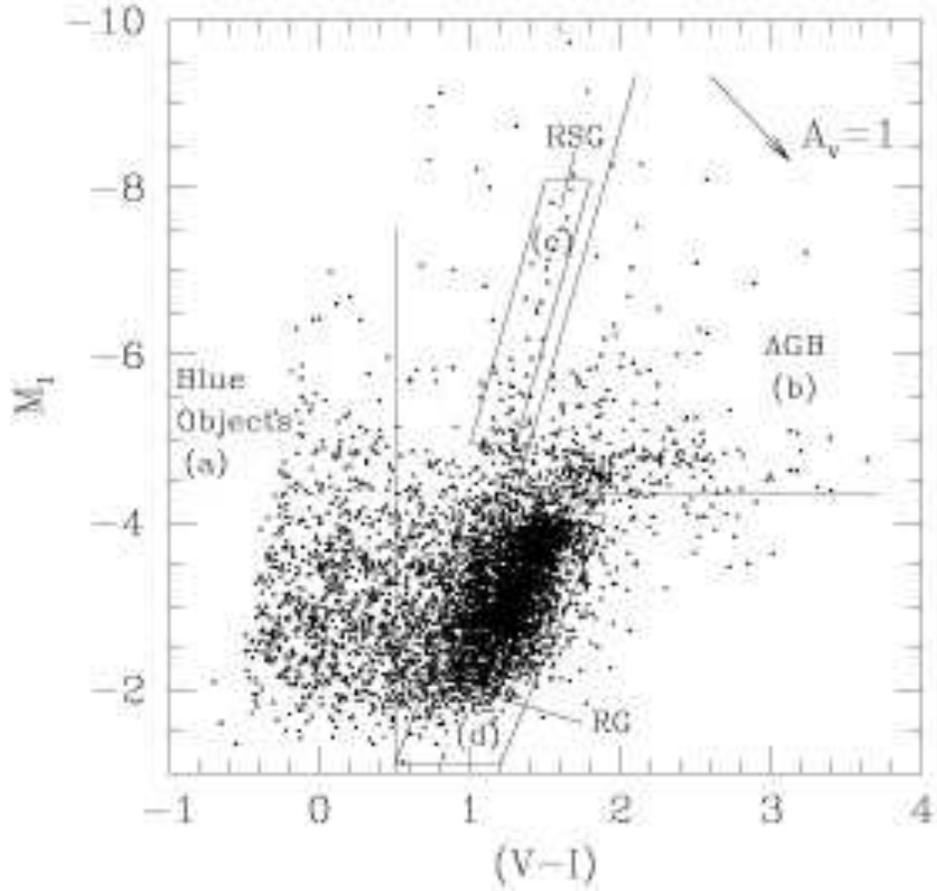}  
\caption{\label{Optical_cmd} Absolute I versus V$-$I CMD from the optical
data. Regions consisting of blue objects (a; left of the vertical line at V$-$I~=~0.5), AGB stars (b), red
supergiants (c), and red giants (d) refer to the the IRAC detections
displayed in Figure \ref{IR_cmd}.}
\end{figure}

As described in \S 2 we created a master photometry list, which
includes detections in V, I, and all four IRAC bands. The optical
properties of the stellar populations have been discussed by other
authors \citep{hod99,rej00,min97} and we will not address them
here. The optical color-magnitude diagram (CMD), shown in Figure
\ref{Optical_cmd} is segregated into sections consisting of blue
objects (a), AGB stars (b), RSGs (c), and red giants (d), enabling us
to ascertain where these stellar types lie in the IR CMD (described
below), as the infrared colors of the objects detected with IRAC are
much less sensitive to effective temperature than those in the
optical. The loci of stellar types were conservatively chosen, with
gaps between them, so that stars will not be misidentified solely due
to photometric errors (though as we discuss below, reddening from dust
absorption can certainly lead to such misidentification).

\begin{figure}
\plotone{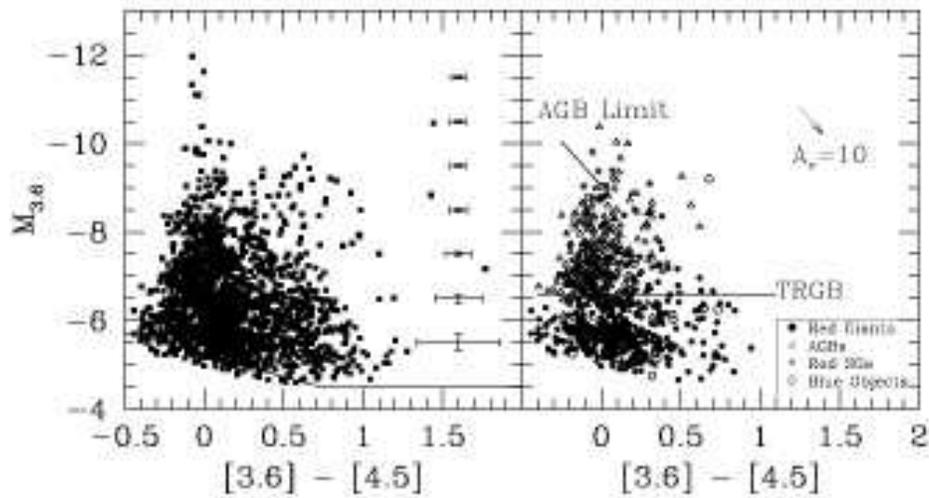}  
\caption{\label{IR_cmd}IRAC absolute 3.6 \micron\ versus [3.6]$-$[4.5]
color-magnitude diagrams for all of the \spitzer\ data (left) and
separated by optical classification for the objects detected in both
the optical and IR (right; see Figure \ref{Optical_cmd}). The left
panel shows 1-$\sigma$ error bars averaged over each magnitude bin an
the 50\% completeness limit (see \S \ref{photometry} for details). The
right panel shows a line connecting the AGB limits for a G0 and M5
stars, a reddening vector for 10 magnitudes of visual extinction, and
the red giant branch tip. Note the paucity of optically detected stars
with very red IR colors.}
\end{figure}

The 3.6 \micron\ absolute magnitude versus [3.6]$-$[4.5] IR CMD is
shown in Figure \ref{IR_cmd}. The basic structure of the IR CMD is a
vertical distribution of stars with [3.6]$-$[4.5] very near zero. This
vertical feature contains all objects where both the 3.6 and 4.5
\micron\ bands sample the Rayleigh-Jeans tail of the Planck
function. Any unreddened object with spectral type earlier than G0
will have [3.6]$-$[4.5] very close to zero, while K0$-$M5 stars become
progressively {\it bluer} to a color of $-$0.25 due to CO absorption at
4.6 \micron\ \citep{jon05}.

A vector showing 10 visual magnitudes of extinction
\citep{ind05,rie85} as well as the AGB limit are included in the IR
CMD. While the reddening vector takes into account the wavelength
dependent extinction from dust, it does not include the circumstellar
emission, which can be significant at longer wavelengths. The AGB
limit was determined by assuming a bolometric correction at 3.6
\micron\ of +3 magnitudes for an M5 star, +2 magnitudes for a G0 star
and M$_{bol}$=$-$7.1 \citep{woo83}, and is depicted in Figure
\ref{IR_cmd} as a line connecting the AGB limit for these two stellar
types.  Above the plotted AGB limit we detect two optically classified
RSGs and one optically classified AGB star. The object optically
classified as an AGB star lies just within the blue boundary of region
(b) and could easily be a slightly reddened RSG. There are also six
objects brighter than, but redward of the AGB limit that are optically
classified as AGB stars. Based on their positions in the optical CMD,
the brightest three of these six objects are likely reddened RSGs,
while the fainter three are consistent with mass-losing AGB stars (see
\S \ref{mass_loss} for a description of the effect of mass-loss on the
[3.6]$-$[4.5] color).

\begin{figure}
\plotone{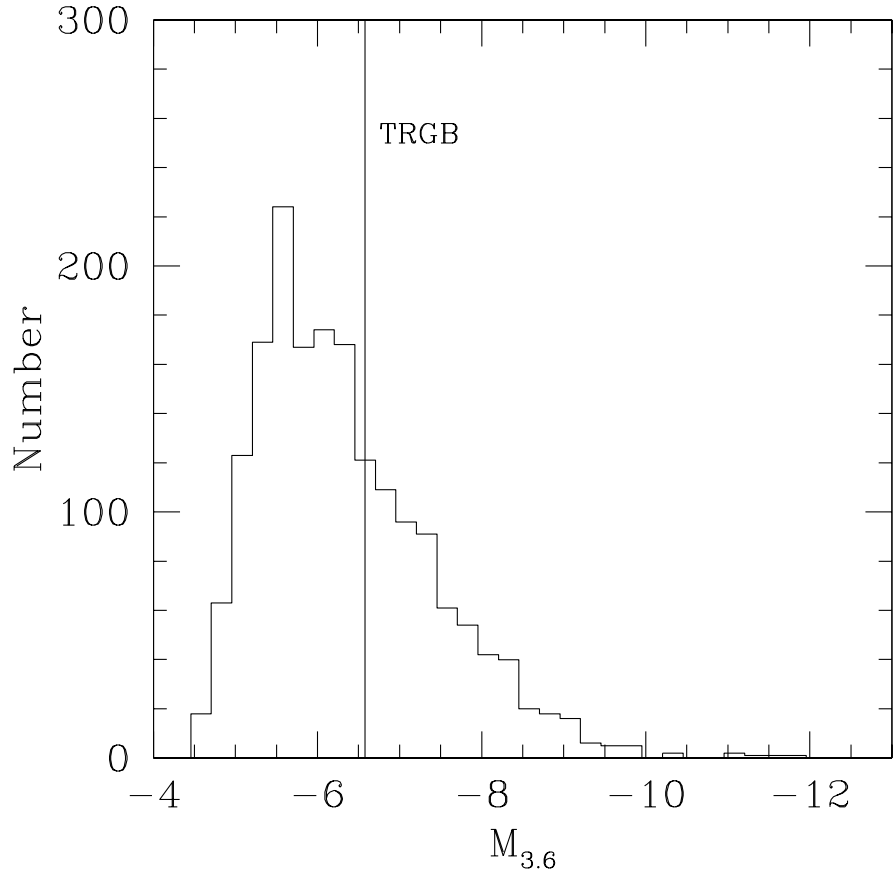}  
\caption{\label{lum_fn}The 3.6 \micron\ luminosity distribution
  (number of stars per 0.25 magnitude bin) for stars detected at both
  3.6 and 4.5 \micron . The TRGB is labeled at M$_{3.6}$~=~$-$6.6.}
\end{figure}

There is considerable overlap in the lower part of the IR CMD between
stellar types. However, the luminosity of the red giant branch tip
(TRGB) does enable us to separate red giants from the other stellar
types. Figure \ref{lum_fn} shows the 3.6 \micron\ luminosity function
for all objects detected at both 3.6 and 4.5 \micron . Based on the
abrupt drop in the 3.6 \micron\ luminosity function, we adopt a 3.6
\micron\ absolute magnitude of the TRGB to be $-$6.6. This value for
the TRGB is slightly brighter than the value adopted by \citet{van05}
for the LMC (L$^\prime$~=~$-$6.4), though given the uncertainty in the
distance modulus to WLM ($\pm$ 0.08 magnitudes) and the bin size of
our luminosity function (0.25 magnitudes) these values agree
reasonably well. For the objects detected at both 3.6 and 4.5 \micron
, 79\% of the total 3.6 \micron\ flux of 46.5 mJy is from stars
brighter than the TRGB, which are predominantly AGB stars.

The right panel of Figure \ref{IR_cmd} shows the IR CMD of objects
detected in V, I, and at 3.6 and 4.5 \micron , separated according to
their optically selected stellar type. The optically detected AGB
stars have typical IR colors of $-$0.25$<$[3.6]$-$[4.5]$<$0.25, with
the exception of a few luminous red sources. The RSGs have a similar
distribution, although no RSGs with red IR colors are detected, as any
reddening would cause a RSG to be misidentified as an AGB star in the
optical. Nearly all of the blue objects detected in the optical and IR
are found between $-$8$<$M$_{3.6}$$<$$-$5 and
$-$0.3$<$[3.6]$-$[4.5]$<$0.5. We also detect a few IR luminous
(M$_{3.6}$~$\sim$~$-$9) unresolved \HII\ regions that are coincident
with strong H$\alpha$ emission. The optically identified red giants
are peaked at M$_{3.6}$~=~$-$6, due to our completeness limit (see
Figure \ref{IR_cmd}), and have a large range in color
$-$0.5$<$[3.6]$-$[4.5]$<$1. As we will discuss in \S \ref{complete},
there are also a large number of optically classified red giants that
lie above the TRGB in the IR, which are consistent with the colors of
reddened AGB stars.

\begin{figure}
\plotone{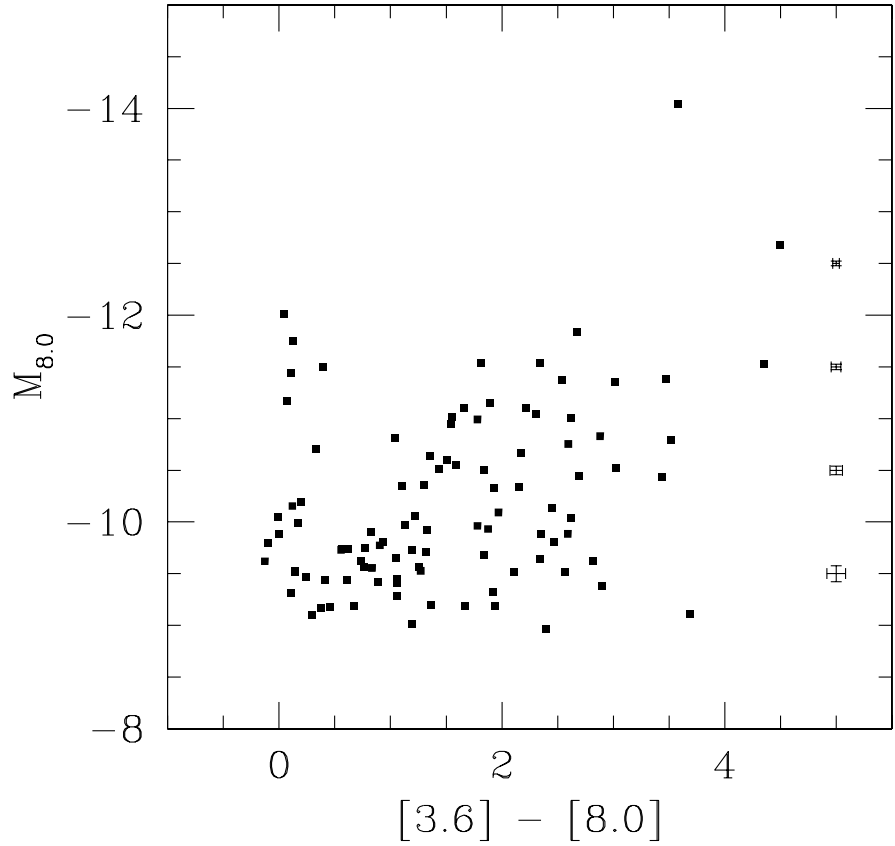}  
\caption{\label{ch4_cmd}IRAC absolute 3.6 \micron\ versus
[3.6]$-$[8.0] color-magnitude diagram. We observe two separate
populations; a narrow distribution with $-$12$<$M$_{3.6}$$<$$-$9 and
[3.6]$-$[8.0]~$\sim$~0 and another, much broader, distribution with
$-$12$<$M$_{3.6}$$<$$-$9 and 1$<$[3.6]$-$[8.0]~$<$4 that are likely
AGBs losing significant mass. Representative 1-$\sigma$ errorbars are
shown at right. }
\end{figure}

Figure \ref{ch4_cmd} is the CMD for all sources detected at both 3.6
and 8.0 \micron . Because of both reduced sensitivity (1-$\sigma$
point source sensitivity of 4.4 $\mu$Jy) and fainter emission from
stellar photospheres, we detect significantly fewer sources at 8.0
\micron\ than in the shorter wavelength bands. However, the longer
wavelength baseline can be useful for separating sources enshrouded in
circumstellar material (such as carbon stars) due to the emission from
dust at 8.0 \micron . In this CMD, apparently two separate populations
are present; a narrow distribution with $-$12$<$M$_{3.6}$$<$$-$9 and
[3.6]$-$[8.0]~$\sim$~0 that are indicative of the brightest AGBs and
RSGs that also have [3.6]$-$[4.5]~$\sim$~0, consistent with normal
AGBs or RSGs not undergoing mass-loss, and another, much broader,
distribution with $-$12$<$M$_{3.6}$$<$$-$9 and 1$<$[3.6]$-$[8.0]~$<$4
that are likely AGBs losing significant mass.

\begin{figure}
\plotone{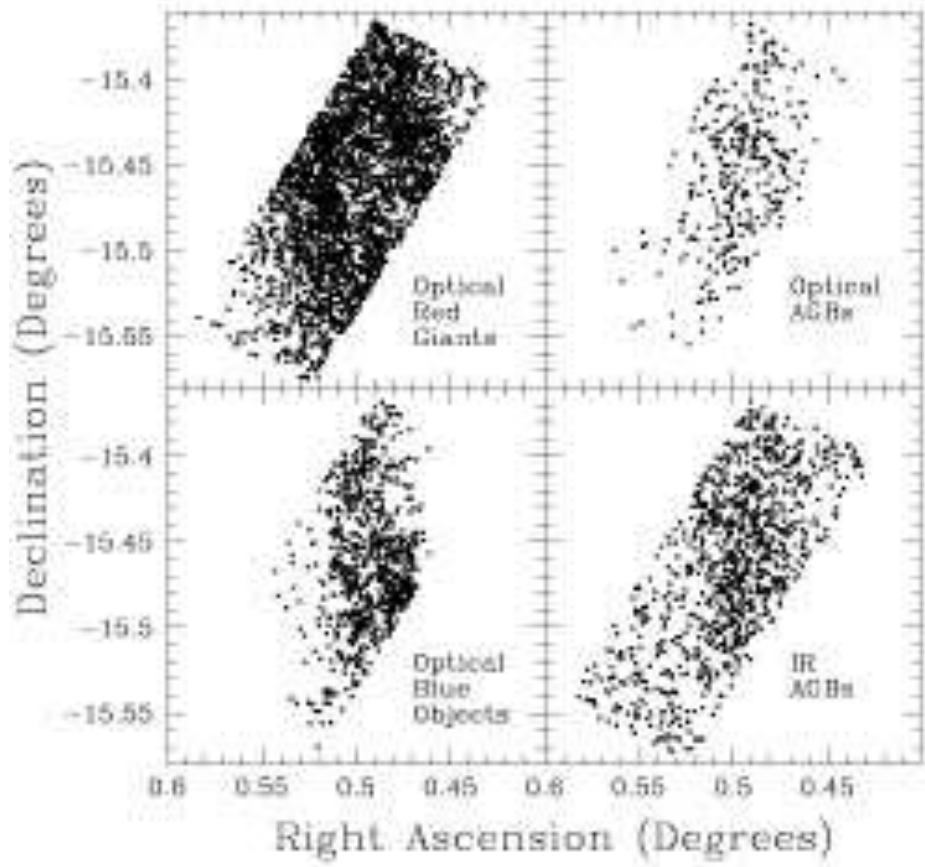}  
\caption{\label{xy}Positions of stars as classified in the optical
(see Figure \ref{Optical_cmd}), except for the bottom right panel, which
shows the positions of AGB stars detected in the IR.}
\end{figure}

In Figure \ref{xy} we show the Right Ascension and Declination of
point sources based on their optical/IR classification. The optically
classified blue objects are strongly concentrated in the regions of
active star formation and the northern arc, while the AGB stars are
much more smoothly distributed. The red giants are also smoothly
distributed over the face of WLM, though this plot shows the effect
that crowding from both blue objects that are very bright in the
optical, and red giants, whose stellar density increases toward the
center of WLM, have on the completeness of fainter sources such as red
giants (e.g., there are `holes' in the detected red giant distribution
toward the center of WLM, particularly at the locations of recent star
formation).

Table \ref{stats} lists many of the detection statistics for our
optical and infrared photometry.  For the region in WLM with coverage
in all four IRAC bands we detect 2855, 2019, 300, and 122 point
sources at 3.6, 4.5, 5.8, and 8.0 \micron , respectively. For the same
region we detect 4989 objects in our matched V and I photometry. We
find 46 objects (38\% of all 8.0 \micron\ detections) that are
detected in all four IRAC bands with no optical counterpart. Of these,
5 have 3.6 and 4.5 \micron\ fluxes consistent with RSG stars (i.e.,
they are above the AGB limit), while the remaining 41 stars could be
either RSGs or AGBs.

In the IR CMD we find a population of bright, red objects to the right
of the main vertical distribution of stars ranging from
M$_{3.6}$~=~$-$10 and [3.6]$-$[4.5]~=~0.3 to M$_{3.6}$~=~$-$7.5 and
M$_{3.6}$$-$[4.5]~=~1.2. Eleven of the 31 objects are detected in the
optical, with seven being AGB stars, two unresolved
\HII\ regions coincident with strong H$\alpha$ emission, and two
objects that are probably foreground giants.  It is likely that many
of the objects not detected in the optical are dust enshrouded AGB
stars, since only the brightest and bluest of these objects are
detected optically and the majority of these are optically classified
as AGBs (see the discussion in \S \ref{mass_loss}).

\section{The AGB Stars}\label{AGB}
Due to the micro-Jy point source sensitivity of IRAC with our
integration time, we should be able to detect the entire population of
the AGB stars in WLM, excluding only those objects that were not
detected due to crowding. However, uniquely determining the true
stellar types of the detected objects can be challenging, as the
spectral energy distributions (SEDs) of these stellar types, with the
exception of unresolved \HII\ regions, are indistinguishable from one
another.

\begin{figure}
\epsscale{0.4}  
\plotone{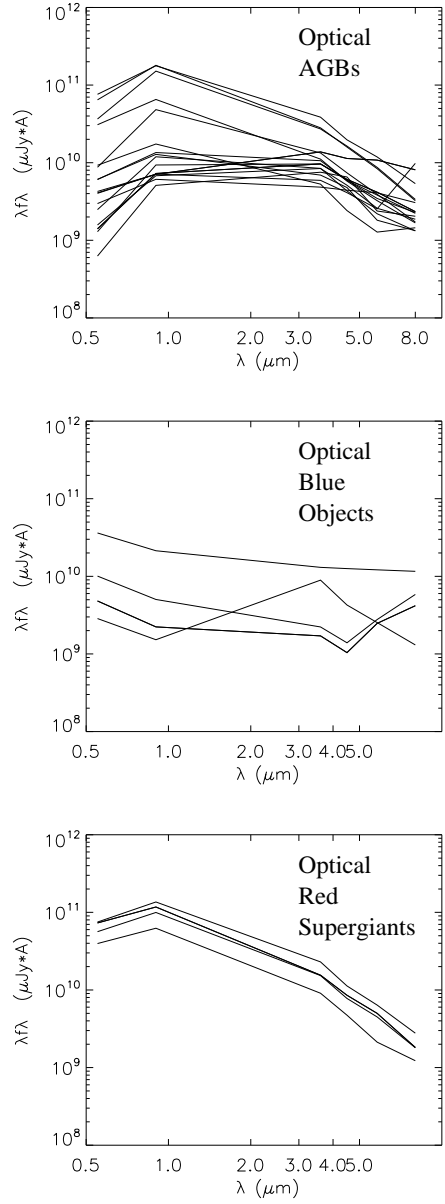}  
\caption{\label{SED} LGGS V and I and IRAC 3-8 \micron\ spectral
energy distributions of (top to bottom) optically identified AGB
stars, blue objects, and red supergiants. Note, in particular, the
similarities between the AGB stars and the RSGs. }
\end{figure}

Figure \ref{SED} shows the IRAC SEDs for the all of the AGB stars,
RSGs, and blue objects detected in all four IRAC bands and the LGGS V
and I images. This plot shows the similarity between the infrared
fluxes of AGBs and RSGs. Although it is challenging to distinguish
between AGB stars and RSGs using IRAC data alone, the combination of
optical and near-IR data provide a powerful method of detecting
candidate AGB stars in galaxies past the immediate vicinity of the
Milky Way, since RSGs are generally easily identified in optical CMDs.

\subsection{Optical Completeness}\label{complete}
One of the goals of our study is to determine the fraction of AGB
stars seen in the IR that were not detected in the optical due to
extinction by circumstellar material. To estimate this fraction, we
make the approximation that all objects above the TRGB in the IR CMD
are AGB stars. While there is contamination above the TRGB from RSGs
and blue objects, the number of these objects detected in the optical
contribute $\lesssim$10\% of the population above the TRGB in the IR.

Of the 691 stars brighter than the TRGB in the IR, 39\% are not
detected in V and I.  There are also 29 objects brighter than the TRGB
in the IR CMD, but lie within region (d) of the optical CMD (the
region surrounding red giants) at positions such that they could have
been reddened from above the TRGB via the reddening vector shown in
Figure \ref{Optical_cmd}. We find it likely that these 29 objects are
AGB stars whose optical fluxes suffer extinction by circumstellar
material.

It is important to consider the effects of photometric scatter as well
as errors due to blending, crowding, and even on our adopted value of
the TRGB on these completeness statistics. By shifting our adopted
value of the TRGB (M$_{3.6}$~=~$-$6.6) up or down by 0.5 magnitudes,
the fraction of optically detected AGB stars changes by only a few
percent.  This is not unexpected though. Most of the AGB stars
detected in the IR but not in the optical are considerably brighter
than the TRGB, because the mass-loss phase that hides AGB stars in the
optical typically takes place at luminosities above the
TRGB. Therefore, even if a number of sources were scattered from above
the TRGB to below, or visa versa, the fraction of optically detected
AGB stars would change very little.  This problem is more significant,
however, for the number of AGB stars misidentified as red giants
because of the large number of optically identified red giants just
below the TRGB. If we shift the adopted value of the TRGB to be 0.5
magnitudes fainter, the fraction of objects we would classify as AGB
stars in the IR, but optically identified as red giants would
increase from 4\% to 8\%. Because the average 3.6 \micron\ photometric
error at the TRGB is only 0.1 magnitudes, it is unlikely that the
fraction of AGBs misidentified as sub-TRGB red giants in the optical
has been effected by more than 1--2\%. 

Further, in the most crowded region of our IRAC images, we detect only
0.04 stars at 3.6 and 4.5 \micron\ per area the size of the 3.6 micron
FWHM disk.  Consequently, we do not anticipate crowding or blending
effects to contribute significantly to the overall photometric errors
of our sources, although it certainly may affect some of them.

In summary, we find that 43\% of the AGB stars in WLM are enshrouded
with enough material to make them either undetectable or misidentified
in the optical CMD, even though in V and I the 50\% completeness level
is nearly three magnitudes below the TRGB. This number of
optically-undetected AGB stars is consistent with other studies
\citep[e.g.,][]{van97,woo98,van06a} that find a significant fraction
of the AGB population (in particular the carbon stars) in an obscured
state. Future investigations of other galaxies from this same
observing program will allow us to determine the dependence of this
fraction on the metallicity of the host system. It should again be
noted that we are ignoring the faint end of the AGB distribution,
i.e., any AGBs fainter than the TRGB.

The completeness fraction discussed above has important implications
for the intermediate age SFH one would derive from galaxies like
WLM. Programs that derive SFHs from optical CMDs
\citep[e.g.,][]{dol02} would tend to over-predict the number of
detected AGB stars, based on the number of main sequence stars, since
many of the AGBs have been reddened beyond the optical detection
limit. Consequently, it is important to apply either corrections to
the number counts in this part of the CMD, or weighting functions to
give more confidence to the number of detected intermediate age main
sequence stars, rather than the AGB stars. As shown here, at least for
WLM, there must be allowance for a factor of 2 in the disparity
between the number of main sequence and AGB stars used to measure the
intermediate age star formation rate. If such disparities are not
accounted for, one will underestimate the true star formation rate
over the past few Gyr.

\subsection{The Carbon Stars}\label{carbon}
\begin{figure}
\epsscale{0.75}  
\plotone{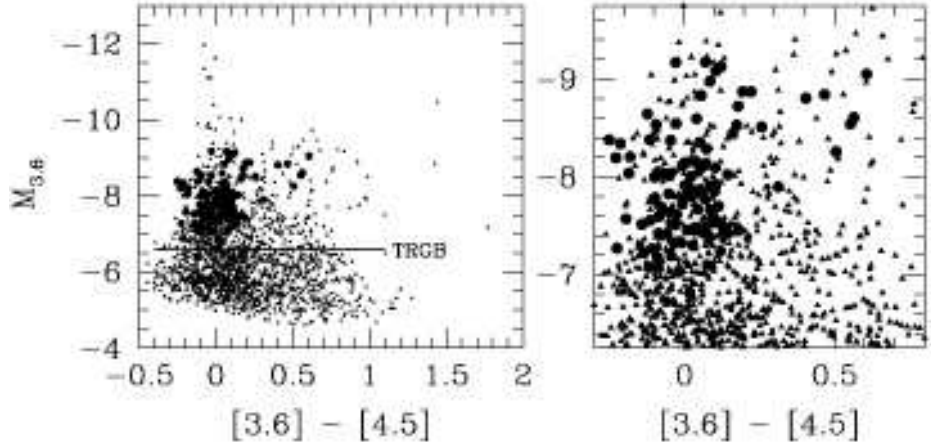}
\caption{\label{bat_cmd} IRAC absolute 3.6 \micron\ versus
[3.6]$-$[4.5] color-magnitude diagram. The small triangles are our
IRAC photometry and the large circles are our measurements of the
carbon stars identified by \citet{bat04}.  The right panel is a
blow-up of the left panel to more easily distinguish the points.}
\end{figure}

\citet{bat04} studied the carbon stars in WLM using
narrow-band optical filters centered on the CN and TiO absorption
features. At 3.6 and 4.5 \micron\ we detect nearly all of the 111
carbon stars detected by \citet{bat04} that are within our coverage
area (our entire IRAC coverage area is within the \citet{bat04}
field). The remaining 11 objects were not detected in the IRAC data
due to crowding, but were detected in the \citet{mas06} optical data.
In Figure \ref{bat_cmd} we plot our IRAC photometry of the carbon
stars identified by \citet{bat04}. The carbon stars span a relatively
wide range of absolute 3.6 \micron\ magnitudes (from 15.75 to 18) and
[3.6]$-$[4.5] colors (from $-$0.3 to 0.6).

The \citet{bat04} photometry is valuable in its ability to distinguish
between carbon-rich and oxygen-rich AGB stars. However, the 111 carbon
stars and 12 M-type AGBs detected in the \citet{bat04} study comprise
only $\sim$18\% of the AGB stars detected in our IRAC data. This is
surprising, given that the broadband portion of their data are shown
to be complete 1 magnitude below the TRGB, and other comparable
studies \citep{alb00,bat03} have estimated completeness near
90\%. There are two explanations that could lead to this
disparity. Either \citet{bat04} have overestimated their completeness
limit, or in this work we have underestimated the contamination above
the TRGB from stellar types other than AGB stars. We know from
comparison with the optical data (\S \ref{photometry}) that at least
10\% of the stars above the TRGB in the IR are not AGB stars. However,
for contamination of other stellar types to be solely responsible for
the difference between these two studies, this fraction would need to
be 90\%, i.e., there would need to be vastly more optically undetected
RSGs, foreground stars, or other objects besides AGBs. This seems
unlikely, considering the relative number of optically detected RSGs
and AGBs shown in Figure \ref{Optical_cmd}, and the very small number
of expected foreground stars.

It is clear from Figure \ref{bat_cmd} that the carbon stars detected
in the narrow-band study have relatively blue [3.6]$-$[4.5] colors,
with only very bright carbon stars being detected with even moderately
red colors. We also detect a number of objects with M$_{3.6}$ between
$-$7 (i.e., the faintest carbon stars detected by the narrow-band
study) and the TRGB. The red AGBs that were not detected by
\citet{bat04} can be explained by circumstellar material reddening
them beyond their detection limits. However, the faint AGBs with blue
[3.6]$-$[4.5] colors that were not detected are somewhat more
perplexing. Because AGB stars cannot become carbon stars until after
thermal pulsing begins (when they are somewhat brighter than the
TRGB), we would expect the region between the TRGB and the faintest
carbon stars to be filled with oxygen-rich AGBs. If the AGBs we detect
between the TRGB and the faintest carbon stars are oxygen-rich AGBs,
they would not be expected to have significant mass-loss rates, and
consequently should be detectable in the optical. One possible
explanation for these non-detections is that a brightness limit in the
\citet{bat04} data was created by the requirement that the combined
broad- and narrow-band photometric errors be small, which would tend
to remove faint, low signal-to-noise detections. There are also a
number of objects with M$_{3.6}$ between $-$7 and the TRGB that have
red [3.6]$-$[4.5] colors. Because AGBs just above the TRGB are not
expected to have large mass-loss rates, the red colors of many of
these objects are likely due to point source blending or photometric
scatter, rather than objects with significant mass-loss.

It is instructive to compare these completeness results with the deep
near-IR imaging of NGC~6822 by \citet{cio05}. They detect a total of
6195 AGB stars in their J and K$_S$ imaging; of which 4684 are
oxygen-rich and 1511 carbon-rich for a carbon star to M-type AGB (C/M)
ratio of 0.32. When including I, J, and K$_S$ they find 2161
oxygen-rich and 500 carbon-rich AGBs for a C/M ratio of 0.23. This is
in comparison with the $\sim$1800 AGB stars (904 carbon stars and
C/M~=~1.0$\pm$0.2) detected in the narrow-band optical study of
NGC~6822 by \citet{let02}.  The fraction of AGBs (both C and M-type)
detected in the optical study of \citet{let02} is somewhat
artificially inflated though, since the field of view of their study
was larger than that in the near-IR study of \citet{cio05}. It is
intriguing to note that while the \citet{let02} optical study detected
significantly fewer AGBs total, the C/M ratio they found was
higher. This is contrary to what we would predict, given the
assumption that carbon stars should be more difficult to detect in the
optical.

It is not clear what the implications of this incompleteness are for
studies of the effect of host galaxy metallicity on the C/M
ratio. Because AGB winds composed of oxygen-rich compounds are less
efficient at absorbing visible photons than those that are carbon-rich
\citep{wal98}, optical studies should be more adept at detecting
M-type rather than carbon-rich AGBs. Consequently, the incompleteness
of the \citet{bat04} study may mean the C/M ratio may be even higher
than the value they derive, C/M~=~12.4, with a large uncertainty
because of the difficulty in distinguishing between M-type AGBs and
foreground Galactic dwarfs. On the other hand, there are a number of
AGBs not detected in the \citet{bat04} study between the TRGB and the
faintest carbon stars, which is where we would expect oxygen-rich AGBs
to reside. This leads us to conclude that many more M-type AGBs
were not detected, lowering the assumed C/M ratio. Shorter-wavelength
near-IR images of WLM (such as in the J, H, and K$_S$ bands) would be
extremely useful in this case, since they could be used to distinguish
between oxygen- and carbon-rich AGB stars and are also less effected
by extinction from circumstellar material than the optical studies.

\subsection{Mass-Loss Rates}\label{mass_loss}
\begin{figure}
\plotone{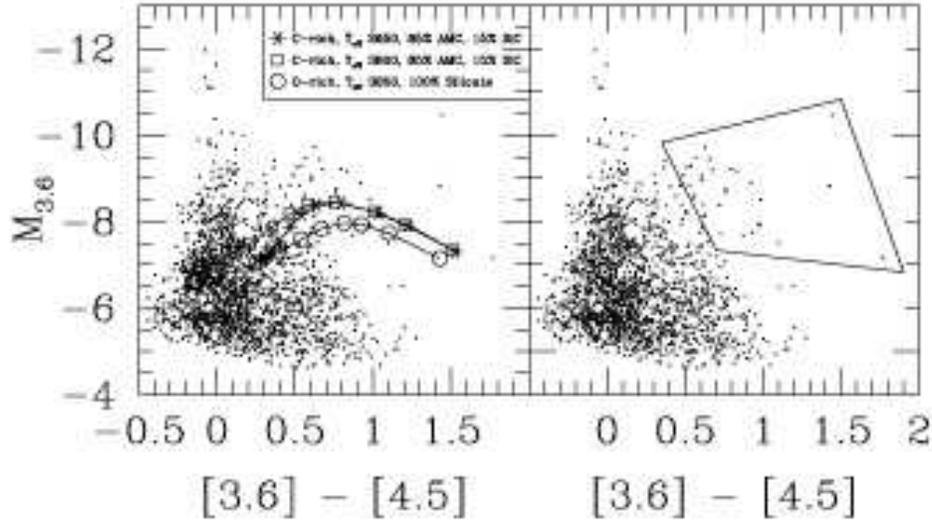}  
\caption{\label{groen} IRAC absolute 3.6 \micron\ versus [3.6]$-$[4.5]
color-magnitude diagrams. The left panel shows tracks for mass-losing
AGBs with luminosity of 3000 L$_\sun$, expansion velocity of 10 km
s$^{-1}$, dust-to-gas ratio of 7.9$\times$10$^{-4}$, and three
different compositions and increasing (from left to right) mass-loss
rates from 6$\times$10$^{-10}$ to 6$\times$10$^{-5}$ for the
carbon-rich tracks and from 6$\times$10$^{-10}$ to 1$\times$10$^{-4}$
\solar\ yr$^{-1}$ for the oxygen-rich track. The box on the right
panel shows the objects selected to derive the conservative total
mass-loss rate discussed in \S \ref{mass_loss}.}
\end{figure}

Using the dust radiative transfer models from \citet{gro06}, we can
estimate wind optical depths ($\tau$) and mass-loss rates (MLRs) for
the AGB stars in our IRAC data. \citet{gro06} presents a number of
models including carbon- and oxygen-rich AGBs and post-AGBs with
varying effective temperatures and wind compositions (including
amorphous carbon (AMC) and silicon carbide (SiC) for the carbon stars
and aluminum oxide (AlOx) and silicates (Si) for oxygen-rich AGBs). In
Figure \ref{groen} we show representative models, displayed as tracks
indicating where an AGB in WLM would move with increasing $\tau$ or
MLR of a given composition.

For the carbon-rich AGBs, these models show very little difference in
the position on the CMD between winds composed of 100\% AMC or 85\%
AMC and 15\% SiC. At large $\tau$ there is also very little difference
between effective temperatures ranging from 2650 K to 3600 K. At small
$\tau$ ($\tau_{\mbox{\scriptsize{11.75 \micron}}}$~$\sim$~10$^{-4}$),
cooler effective temperatures correspond to the same M$_{3.6}$
magnitude but redder [3.6]$-$[4.5] colors. If the main vertical
distribution in our IR CMD is composed of carbon stars with negligible
MLRs, the majority have effective temperatures slightly cooler than
3600 K. It is interesting to note that while some of the spread in
color of the vertical distribution is certainly due to increasing
photometric scatter with increasing M$_{3.6}$ magnitude, this spread
is also consistent with the range of modeled effective temperatures.

For the oxygen-rich AGBs at a given effective temperature, varying the
composition of the wind from 100\% AlOx, to 60\% Si and 40\% AlOx, and
to 100\% Si yield tracks roughly the same shape. However, increasing
the Si fraction shifts the tracks redward. Changing the spectral type
for a given wind composition from M0 to M6 to M10 produces brighter
M$_{3.6}$ magnitudes and redder [3.6]$-$[4.5] colors.

Because of the similarities between the tracks of oxygen- and
carbon-rich AGBs, we cannot distinguish between them in our
CMD. However, we can still investigate the MLRs of AGB stars by
comparing the distribution of our sources to the various models. The
[3.6]$-$[4.5] color of AGB stars corresponds to a unique $\tau$, which
in turn corresponds to a MLR, scaled by the square root of the stellar
luminosity and by the inverse of the dust-to-gas ratio ($\psi$) for a
given stellar/wind composition, as prescribed by \citet{gro06}. We
calculated the total MLRs by binning the AGBs according to the modeled
optical depths and applying the appropriately scaled MLR, given the
stellar color and luminosity. \citet{dol00} estimates the
metallicity of WLM 1-2.5 Gyr ago (an appropriate timescale for the
formation of the current AGB population) to be [Fe/H]~=~$-$1.13, while
the current metallicity of WLM, as measured by both the nebular oxygen
abundance \citep{lee06} and blue supergiants \citep{bre06} is
[O/H]~=~$-$0.8. These upper and lower limits give us a reasonable
range of values for the expected metallicities of the AGB population,
and also allow us to estimate an uncertainty in the calculated
MLRs. Assuming the dust-to-gas ratio scales as
$\psi$~=~$\psi_{\sun}$~10$^{-[Fe/H]}$ and $\psi_{\sun}$~=~0.005
\citep{van05}, we adopt $\psi_{WLM}$~=~3.7-7.9$\times10^{-4}$ .

If the entire AGB population is composed of carbon-rich AGBs with
T$_{eff}$~=~3600 K and wind composition of 85\% AMC and 15\% SiC we
derive a total MLR of \ml~=~1.1-2.4$\times$10$^{-3}$ \solaryr .  If
instead we assume a population with the same properties except
T$_{eff}$~=~2650 K, we find a similar total MLR of
\ml~=~0.9-1.9$\times$10$^{-3}$ \solaryr . In both of these cases we
are not including the contribution from sources bluer than the model
colors; however, these objects contribute a negligible amount to the
total MLR. Ninety percent of the total MLR from both models are from
sources with [3.6]$-$[4.5]~$>$~0.5, and over 50\% of the mass-loss is
from sources with [3.6]$-$[4.5]~$>$~1.0. The total MLRs are a factor
of 3.3 higher if the population is composed of oxygen-rich AGBs with
60\% Silicate and 40\% AlOx winds, and a factor of 2.9 higher for
oxygen-rich AGBs with 100\% silicate winds.

We are particularly interested in the sources redward of the main
stellar distribution (shown in the box in the right panel of Figure
\ref{groen}). These sources are well modeled by both oxygen-rich AGBs
with $\tau_{\mbox{\scriptsize{11.75 \micron}}}$~=~3.0 and carbon-rich
AGBs with $\tau_{\mbox{\scriptsize{11.75 \micron}}}$~=~0.54. The only
models ruled out by these data are an entire population of oxygen-rich
AGBs with a wind composed of 100\% aluminum oxide (at any T$_{eff}$),
since even at the largest modeled $\tau$
($\tau_{\mbox{\scriptsize{11.75 \micron}}}$~=~18.5) these objects do
not become red enough to reproduce the observed distribution. It is
appropriate to calculate a more conservative MLR, based solely on
these objects, since many sources that are consistent with having
moderate MLRs may have artificially red colors due to photometric
scatter and/or point source blending. The MLRs calculated using only
these objects are 0.8-1.6$\times$10$^{-3}$ \solaryr\ for carbon stars
with T$_{eff}$~=~3600 K and 0.7-1.4$\times$10$^{-3}$ \solaryr\ for
T$_{eff}$~=~2600 K. This box was chosen somewhat arbitrarily, however,
moving the blue boundary of the box to include AGBs with moderate
[3.6]$-$[4.5] colors changes the total MLR by less than 10\%, since
the overall mass-loss is dominated by the handful of very red sources.

Because the mass-loss rates we derive are so strongly dependent on the
few reddest sources in this galaxy, it is imperative that we be
confident that these sources are real and have not been created by
photometric errors or the faulty merging of photometry from different
filters. Of the 33 sources in the `conservative' box in Figure
\ref{groen}, all are also detected at 5.8 \micron\ and all but one are
detected at 8.0 \micron . Additionally, the SEDs of these objects show
no unexpected features, such as those that would be obvious if the the
flux in one filter from a red giant was merged with the other three
filters from an AGB star. Some caution is in order here, however,
because the total mass-loss in WLM is dominated by only a few sources
and the super-wind phase of mass-loss for an individual AGB star is
very short ($\sim$ 5$\times$10$^4$ years; \citealt{vas93}) compared to
evolutionary timescale for the entire stellar
population. Consequently, the mass-loss measured at the present epoch
may not be representative of the average mass-loss over an extended
period in the galaxy.

We compare these mass-loss values with the models from \citet{ken94}
which estimate the return of material into the ISM from a single-age
stellar population. They find 26-46\% of the initial stellar mass is
returned into the ISM via supernovae and AGB winds, depending on the
assumed IMF, and show most of this material is returned within the
first Gyr. \citet{dol05} finds the star formation rate over the last
Gyr in one HST/WFPC2 field in WLM to be 1$\times$10$^{-3}$ \solaryr
. Within this WFPC2 field we detect seven AGBs in the IR that
contribute a total MLR of 1.8-3.8$\times$10$^{-4}$ \solaryr
. Consequently, our conservative MLR represents a 18-38\% return of
mass into the ISM.  \citet{van05} showed that about half of the total
mass-loss an AGB star experiences occurs in the dust-enshrouded phase.
Because the AGBs we use to calculate this rate all have red
[3.6]$-$[4.5] colors that are indicative of the dust-enshrouded phase,
to estimate the total mass-loss from these objects it is reasonable to
adjust this value upward by a factor of two, to 36-76\%. These values
are probably overestimated, given that they are higher than the
\citet{ken94} prediction and we are not measuring the contribution of
material from supernovae and massive AGBs that have already died. Some
possible explanations for this overestimation are the lower limit for
our assumed dust-to-gas ratio may be too low, which would increase the
derived MLR for a given wind optical depth, and also any `burstiness'
in the true star formation rate, which could either raise or lower the
calculated fraction being returned to the ISM compared to the average
value over the past Gyr. Given the fluctuations observed in the star
formation histories of dIs, the factor of two agreement is probably
quite reasonable.

\begin{figure}
\plotone{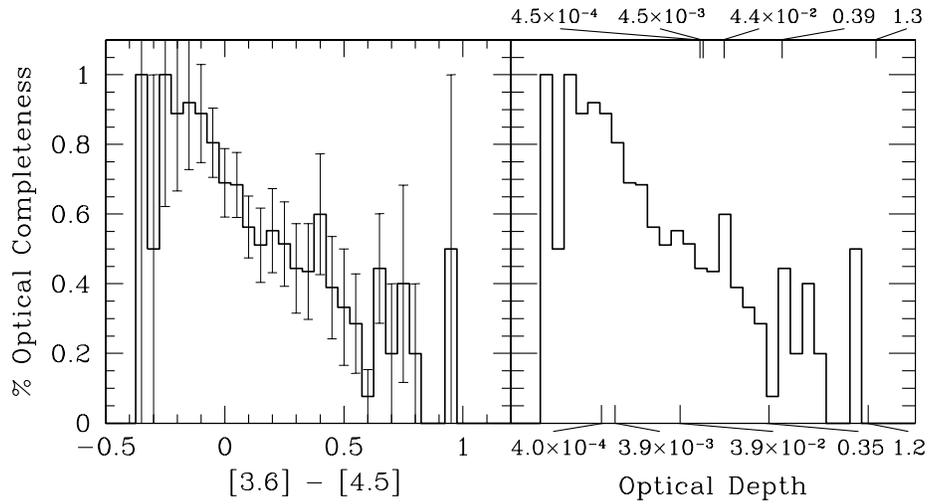}  
\caption{\label{comp_frac} The fraction of objects brighter than the
TRGB at 3.6 \micron\ that are detected in the optical as a function of
[3.6]$-$[4.5] color (left panel) and wind optical depth (right panel)
for a carbon-rich AGB with 85\% AMC and 15\% SiC wind and effective
temperatures of 2650 K (top axis) and 3600 K (bottom axis). The
errorbars were determined by taking the square root of the number of
optically detected objects divided by the total number detected at 3.6
and 4.5 \micron . The trend of decreasing optical completeness with
increasing optical depth (and MLR) is clearly shown.}
\end{figure}

In Figure \ref{comp_frac} we show a histogram of the fraction of
objects brighter than the TRGB that were detected in the optical as a
function of [3.6]$-$[4.5] color (left panel) and optical depth for
carbon-rich AGBs with 85\% AMC and 15\% SiC wind and two different
effective temperatures (right panel). This figure clearly shows the
trend of decreasing optical completeness with increasing MLR, which
supports our conclusion that the AGB stars misidentified or not
detected in the optical have been reddened by circumstellar material.

\citet{fro98} report that there are very few optically detected carbon
stars in the Large Magellanic Cloud (LMC) with MLRs above 10$^{-6}$
\solaryr, and none above 10$^{-5}$ \solaryr. They adopt a value of
\ml~=~5$\times$10$^{-6}$ \solaryr\ as the critical value above which
no carbon star will be detected optically. We detect no AGBs optically
with [3.6]$-$[4.5]~$>$~1.0 ($\tau$~=~1.3,
\ml~$\sim$~5$\times$10$^{-5}$ \solaryr\ assuming L~=~3000 L$_{\sun}$)
and only one with [3.6]$-$[4.5]~$>$~0.85 ($\tau$~=~0.7,
\ml~$\sim$~2$\times$10$^{-5}$ \solaryr ) out of the 14 AGBs detected
in the IR. Given the difference in metallicity of the LMC and WLM
($\sim$ 0.7 dex) and its effect on the derived MLR, these values are
in agreement with the \citet{fro98} result.

\begin{figure}
\epsscale{0.5}  
\plotone{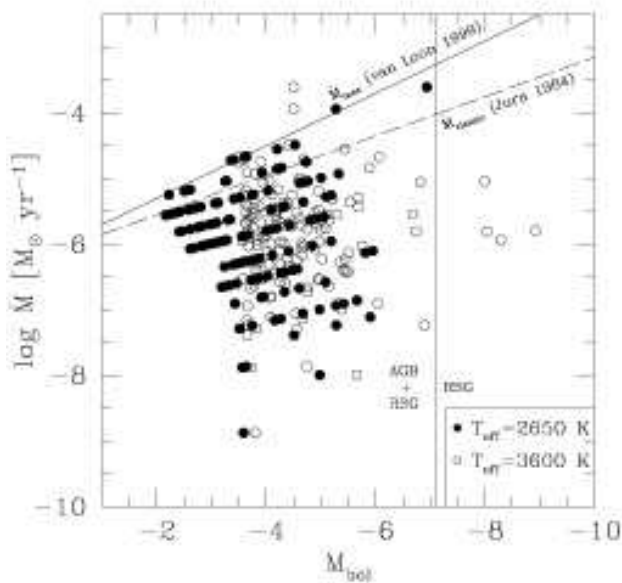}  
\caption{\label{Mbol} Mass-loss rates versus bolometric magnitude for
all objects brighter than the TRGB at 3.6 \micron , assuming
dust-to-gas ratio of 7.9$\times$10$^{-4}$, a wind composition of 85\%
AMC + 15\% SiC and effective temperatures of T$_{eff}$~=~2650
({\it filled} circles) and T$_{eff}$~=~3600 ({\it open} circles),
following \citet{van99}.  The bottom dashed line is the classical
single-scattering mass-loss limit \citep{jur84} and the top solid line
is the empirical maximum mass-loss limit suggested for the LMC by
\citet{van99}. The AGB limit is shown (vertical line) at
M$_{bol}$=$-$7.1. See \S \ref{mass_loss} for discussion.}
\end{figure}

Figure \ref{Mbol} shows the MLR versus bolometric luminosity for all
of the objects brighter than the TRGB in our IR CMD. Filled circles
represent the MLRs and luminosities assuming all of the AGBs are
carbon-rich with dust-to-gas ratios of 7.9$\times$10$^{-4}$, winds
composed of 85\% AMC and 15\% SiC, and T$_{eff}$~=~2650 K. Open
circles are for the same composition with T$_{eff}$~=~3600 K. We also
show the classical single-scattering mass-loss limit (bottom dashed
line) and the empirical maximum mass-loss limit found for the LMC (top
solid line) as plotted in \citet{van99}. Because we do not know the
effective stellar temperature a priori, some MLRs can be overestimated
in the case of the T$_{eff}$~=~3600 K template or underestimated for
T$_{eff}$~=~2650 K, since there is a degeneracy in our CMDs between
stars with cool effective temperatures and stars with warmer effective
temperatures but significant MLRs. Consequently, some caution should
be taken in strictly interpreting this plot. The dust-to-gas ratio
assumed for this plot is the upper limit of the two values we consider
in this paper. By instead assuming the lower limit, all mass-loss
rates would be scaled upward by approximately a factor of two.

Despite these uncertainties, there is a remarkable consistency between
the AGB populations of WLM and both the LMC and SMC
\citet{van99,van06b}.  Under the assumption of T$_{eff}$~=~2650 K, the
empirical maximum MLR found in the LMC by \citet{van99} is in
excellent agreement with the AGBs in WLM. We also detect many AGBs
that are consistent with $\dot{M}$~=~10$^{-7}$ \solaryr\ and
M$_{bol}$~=~$-$4. AGBs with these properties were also seen in the LMC
\citet{van99}, though relatively few due to sensitivity
considerations. \citet{gai87} show that these objects could not exist
if the wind-driving mechanism was purely by the classical method of
radiation pressure accelerating dust grains whose momenta are coupled
to circumstellar gas \citep{geh71}. These observations support those
of \citet{van99}, suggesting another mechanism may be at work.

\subsection{Luminosity Contribution From AGB Stars}\label{luminosity}
One of the reasons for observing galaxies in the IR is to obtain
accurate measurements of their stellar masses. When using luminosity
measurements to estimate stellar masses at any wavelength, it is
important to understand the stellar types contributing to that
luminosity. In the optical, the luminosity can be dominated by
massive, young stars that describe only the very recent star formation
rate, while in the near-IR the luminosity is from old red giants (in
the short-wavelength near-IR) or intermediate age AGB stars (in the
IRAC bands). Nearby galaxies offer us the opportunity to determine
precisely which stars contribute to a galaxy's total luminosity as a
function of wavelength.  As we note in \S \ref{AGB}, for the point
sources detected at 3.6 and 4.5 \micron , 79\% of the total flux is
from AGB stars, in contrast with optical and shorter
wavelength near-IR studies.

\citet{lee06} used 4.5 \micron\ luminosities to estimate total stellar
masses of nearby dwarf galaxies. We recover 83\% of the total 4.5
\micron\ flux \citet{lee06} find for WLM in point sources we detect at
both 3.6 and 4.5 \micron . Although the 4.5 \micron\ luminosity of WLM
is dominated by the light from AGB stars, because the evolutionary
phase of AGB star production is relatively long, the 4.5 \micron\
luminosity is a fairly robust measure of the stellar mass (i.e.,
short-term variations in the SFH are averaged out). However, for
smaller galaxies or galaxies with unusual SFHs,
the 4.5 \micron\ luminosity may be less directly connected to the
stellar mass.

\citet{van05} provided a method of estimating the masses of stellar
clusters based on the number of stars brighter than the TRGB in the
L$^\prime$ band (3.76 \micron ) and the age of the cluster. Performing
this calculation for WLM is somewhat difficult because, unlike stellar
clusters, its stellar population is not single aged. However, by
adopting an age of 2 Gyr for the stellar population \citep[i.e., the
age of the recent star forming event that formed a significant
fraction of the stars;][]{dol00}, we arrive at a total stellar mass of
1.1$\times$10$^7$ \solar . Because of the range in age of the stellar
population and our incomplete sky coverage this value is probably
uncertain by at least 50\%; however, it is reasonably consistent with
the \citet{lee06} value of 1.8$\times$10$^7$ \solar .


\section{Results and Conclusions}\label{conclusions}
We have presented \spitzer /IRAC imaging at 3.6 and 4.5 \micron .  V
and I optical data from the LGGS \citep{mas06}
were combined with these observations to help determine the stellar
types of objects detected in the IR. We find that above the TRGB at
3.6 \micron , 39\% of the objects were undetected in the optical and
43\% were undetected or misidentified, even though the \citet{mas06}
data are 50\% complete 3 magnitudes below the TRGB. It is likely that
many of these objects were not detected in the optical due to dust
production in the winds of the AGB stars, which efficiently absorb
visible photons while IR radiation is relatively unaffected.

Comparing our photometry with the narrow-band optical study of WLM
\citep{bat04}, we detect nearly all of the objects they classify as
carbon stars with a few exceptions, mostly due to crowding
effects. However, of the 691 stars we detect and classify as AGBs only
18\% were detected by the narrow-band study. It is unclear what effect
this may have on the assumed C/M ratio, since there is evidence that
both mass-losing carbon stars and M-type AGBs both escaped detection
in the \citet{bat04} study.

Additionally, we detect a significant population of objects with
positions on the IR CMD consistent with mass-losing AGBs. For these
objects we derive a conservative total MLR for the galaxy of
0.7-1.6$\times$10$^{-3}$ \solar\ yr$^{-1}$, assuming the entire
population is composed of carbon stars. If we include all of our 3.6
and 4.5 \micron\ detections, we find a total MLR of up to
0.9-2.4$\times$10$^{-3}$ \solar\ yr$^{-1}$, which is likely an
overestimate due to the inclusion of objects photometrically scattered
to their relatively red positions in the IR CMD.

For the region of the galaxy covered by the HST/WFPC2 study of
\citet{dol00}, we find 36-76\% of the initial mass in stars formed
over the last Gyr is currently being returned into the ISM via
mass-loss from AGB stars, assuming half of the total mass lost by an
AGB star occurs in the dust enshrouded phase. This value is somewhat
higher than expected, given a total of 26-46\% is estimated to be
returned from the entire stellar population \citep{ken94}, and our
data are not sensitive to the contribution from supernovae and AGBs
that have already died. The prescription we use may systematically
overestimate the MLRs.

We find the distribution of MLRs and bolometric luminosities of AGBs
and RSGs are very similar to those in the LMC and SMC
\citep{van99,van06b} and the empirically derived maximum MLR of the LMC
\citep{van99} is in excellent agreement with our data.

Finally, we show that the AGB stars, which trace the intermediate age
stellar population, provide the dominant contribution to the
luminosity in the IRAC bands, while studies at shorter wavelengths can
preferentially detect either much younger sources (in the optical) or
older age populations (short-wavelength near-IR).


\acknowledgements We thank Martin Groenewegen for his comments on a
previous version of this manuscript and for providing his models of
mass-losing AGBs prior to publication. We thank Jacco van Loon and
Maria-Rosa Cioni for their valuable comments and suggestions, which
contributed significantly to the presentation of these data.  We also
thank the anonymous referee for their prompt and careful reading of
the manuscript and their valuable comments.  E.~D.~S. acknowledges
partial support from a NASA LTSARP grant NAG 5-9221 and the University
of Minnesota. This work is based in part on observations made with the
Spitzer Space Telescope, which is operated by the Jet Propulsion
Laboratory, California Institute of Technology under NASA contract
1407. Support for this work was provided by NASA through Contract
Numbers 1256406 and 1215746 issued by JPL/Caltech to the University of
Minnesota. This research has made use of NASA's Astrophysics Data
System Bibliographic Services and the NASA/IPAC Extragalactic Database
(NED) which is operated by the Jet Propulsion Laboratory, California
Institute of Technology, under contract with the National Aeronautics
and Space Administration.


\clearpage


\clearpage
\begin{deluxetable}{lcc}
\tablecaption{Basic Properties of WLM}
\tablewidth{0pt}
\tablehead{
\colhead{Quantity} & \colhead{Value} & \colhead{Reference}}
\startdata
Right Ascension, $\alpha$(2000) & 00 01 57.8 & 1\\
Declination, $\delta$(2000) & $-$15 27 51 & 1\\
Heliocentric velocity, V$_\sun$ (km s$^{-1}$) & $-$130 km s$^{-1}$ & 5\\
Distance, D (Mpc) & 0.92 $\pm$ 0.04 & 6\\
Morphological Type & Ir IV-V & 4\\
12~+~log(O/H) & 7.83 & 6 \\
Total \ion{H}{1} mass (M$_\sun$) & 5.3$\times10^7$ & 3\\
Inclination angle (degrees) & 69 & 2\\
Position angle (degrees) & 181 & 5\\
Rotational velocity (km s$^{-1}$) & 38 & 5\\
Conversion factor (pc/arcmin) & 276 & 5\\
\enddata
\label{basic}
\tablerefs{(1) \citet{gal75}. (2) \citet{abl77}.  (3) \citet{huc81}. (4) 
\citet{van94}. (5) \citet{jac04}. (6) \citet{lee06}.}
\end{deluxetable}   
\clearpage


\clearpage
\begin{deluxetable}{lc}
\tablecaption{\label{stats} Detection Statistics}
\tablewidth{0pt}
\tablehead{
\colhead{\hspace{2cm}Total Point Source Detections in All Wavelengths }}
\startdata
Filter & Number \\
\hline
Both V and I & 4989     \\
3.6 \micron & 2855 \\
4.5 \micron & 2019 \\
5.8 \micron & 300  \\
8.0 \micron & 122  \\
& \\
\multicolumn{2}{c}{Detections in All Four IRAC bands, But Not V and I} \\
\hline
Object type & Number \\
\hline
Total & 46 \\
RSG (Above the AGB limit)& 5 \\ 
AGB/RSG (Above the TRGB, below the AGB limit)& 41 \\
& \\
\multicolumn{2}{c}{3.6 \micron\ Point Source Flux (Total Flux = 46.5 mJy)} \\
\hline
Object type & Fraction \\
\hline
Brighter than the TRGB & 79\% \\
Fainter than the TRGB & 21\% \\
& \\
\multicolumn{2}{c}{Optical Detection Fractions of IR Identified AGBs} \\
\hline
Filter & Fraction \\
\hline
V and I & 61\% \\
Detected but misidentified in V and I & 4\% \\
Narrow-band optical & 16\% \\
\enddata
\end{deluxetable}   
\clearpage


\end{document}